\documentclass[aps,prb,twocolumn,superscriptaddress,showpacs,showkeys]{revtex4-2}

\usepackage{color}
\usepackage{graphicx}
\usepackage{dcolumn}
\usepackage{bm}
\usepackage{float}
\usepackage[mathlines]{lineno}
\usepackage{tabularx}
\usepackage[colorlinks,linkcolor=blue,anchorcolor=blue,citecolor=blue,urlcolor=blue,hyperindex,CJKbookmarks]{hyperref}
\usepackage{footnote}
\usepackage{amsmath}
\usepackage{txfonts}
\usepackage{mathptmx}
\usepackage{ragged2e}
\usepackage{booktabs,makecell, multirow, tabularx}
\usepackage{multirow}
\usepackage{braket}
\usepackage{gensymb}
\usepackage{array}
\usepackage{makecell} 
\usepackage{booktabs} 
\usepackage{svg}
\usepackage{siunitx}
\usepackage{xcolor}

\newcolumntype{L}[1]{>{\raggedright\let\newline\\\arraybackslash\hspace{0pt}}m{#1}}
\newcolumntype{C}[1]{>{\centering\let\newline\\\arraybackslash\hspace{0pt}}m{#1}}
\newcolumntype{R}[1]{>{\raggedleft\let\newline\\\arraybackslash\hspace{0pt}}m{#1}}

\begin{document}

\title{Layer-resolved Electronic Structure and Correlation of Low-$n$ Square-planar Nickelates: A DFT+DMFT Prediction of Superconducting Candidates}

\author{Jian-Hong She}
\affiliation{School of Physics and Key Laboratory of Quantum State Construction and Manipulation (Ministry of Education), Renmin University of China, Beijing 100872, China}

\author{Rong-Qiang He}\email{rqhe@ruc.edu.cn}
\affiliation{School of Physics and Key Laboratory of Quantum State Construction and Manipulation (Ministry of Education), Renmin University of China, Beijing 100872, China}

\author{Zhong-Yi Lu}\email{zlu@ruc.edu.cn}
\affiliation{School of Physics and Key Laboratory of Quantum State Construction and Manipulation (Ministry of Education), Renmin University of China, Beijing 100872, China}
\affiliation{Hefei National Laboratory, Hefei 230088, China}

\date{\today}

\begin{abstract}
Multi-layer square-planar nickelates provide a rare platform in which the nominal Ni valence, dimensionality, and layer-resolved electronic structure can be tuned within the same structural family. Recent experiments have found superconductivity in $n=4$--8 $R_{n+1}Ni_nO_{2n+2}$ compounds, with the highest $T_c$ near $n=6$, whereas the more heavily hole-doped $n=3$ member remains nonsuperconducting. Here we propose spacer-layer Cl doping as a route to convert low-$n$ nickelates into superconducting candidates. Compared with changing the layer number $n$, Cl substitution on the spacer-layer oxygen sites offers a chemically natural way to continuously tune the Ni valence while leaving the NiO$_2$ planes largely intact; the lower-$n$ compounds may also be more accessible for synthesis. Using density functional theory combined with dynamical mean-field theory, we show that electron-compensated $n=2$ and $n=3$ La-based nickelates, targeted to the nominal Ni valence of superconducting $n=6$, develop Ni-$d$ correlations comparable to those of superconducting higher-$n$ compounds while preserving the characteristic low-energy Ni-$d$ electronic structure. These results suggest spacer-layer Cl doping as a promising strategy for designing low-$n$ square-planar nickelate superconductors.
\end{abstract}


\maketitle

\section{Introduction}
The discovery of superconductivity in infinite-layer nickelates has revived the long-standing search for cuprate analogues based on square-planar NiO$_2$ layers \cite{Li2019_nature}. The original motivation is simple: Ni$^{1+}$ in $R$NiO$_2$ has the same nominal $3d^9$ configuration as Cu$^{2+}$ in parent cuprates, and early theoretical works proposed reduced nickelates as promising platforms for cuprate-like superconductivity \cite{Bednorz1986_zpb,Keimer2015_nature,Anisimov1999_prb}. Superconductivity was first reported in Sr-doped NdNiO$_2$ thin films \cite{Li2019_nature}, followed by superconducting domes in Nd$_{1-x}$Sr$_x$NiO$_2$ \cite{Li2020_prl,Zeng2020_prl}, Pr$_{1-x}$Sr$_x$NiO$_2$ \cite{Osada2020_prm}, La$_{1-x}$Sr$_x$NiO$_2$ \cite{Osada2021_adma}, and La$_{1-x}$Ca$_x$NiO$_2$ \cite{Zeng2022_sciadv}. These results established square-planar nickelates as a new family of unconventional superconductors, as summarized in recent reviews \cite{Wang2024_arcmp,Wang2025_nsrreview}. At the same time, both experiments and theory have shown that nickelates are not simple copies of cuprates. Related DFT+DMFT work on designed Ruddlesden-Popper cuprates also suggests that nickelate-like two-orbital correlations can be obstructed by the charge-transfer nature of cuprates \cite{She2026_prb_cu}. Compared with cuprates, infinite-layer nickelates have weaker O-$2p$/Ni-$3d$ hybridization, larger charge-transfer energy, rare-earth-derived metallic bands, and multiorbital low-energy electronic structures \cite{Lee2004_prb,Hepting2020_natmater,Gu2020_commphys,Ding2024_nsr_il,Si2024_prresearch}. The relevant control parameters for nickelate superconductivity therefore remain under active debate.

Recent progress has greatly expanded the nickelate materials landscape beyond infinite-layer $R$NiO$_2$. High-pressure superconductivity was reported in bilayer La$_3$Ni$_2$O$_7$ with $T_c$ approaching 80 K \cite{Sun2023_nature,Zhang2024_natphys,Wang2024_prx_l327}, and later in trilayer La$_4$Ni$_3$O$_{10}$ \cite{Zhang2025_prx}. The bilayer discovery has triggered extensive studies of orbital-dependent correlations, partial-gap behavior, magnetic excitations, and interlayer-coupling-driven pairing mechanisms \cite{Yang2024_natcomm_l327,Liu2024_natcomm_l327_optics,Chen2024_natcomm_l327_rixs,Jiang2024_cpl_l327,Sakakibara2024_prl_l327,Lu2024_prl_l327,Luo2024_npjqm_l327,Ouyang2024_prb_hund_l327,Tian2024_prb_swave_l327,Chen2024_prb_hole_l327}. Related DFT+DMFT studies have also emphasized Hund correlations and non-Fermi-liquid behavior in La$_4$Ni$_3$O$_{10}$ and in the monolayer-trilayer phase of La$_3$Ni$_2$O$_7$ \cite{Wang2024_prb_l4310,Ouyang2025_prb_1313}. In the reduced square-planar series $R_{n+1}$Ni$_n$O$_{2n+2}$, superconductivity was first found in the quintuple-layer compound Nd$_6$Ni$_5$O$_{12}$ \cite{Pan2022_natmater}. More recently, a superconducting phase diagram was established for multi-layer Nd$_{n+1}$Ni$_n$O$_{2n+2}$, where superconducting signatures appear for $n=4$--8, with the highest $T_c$ near $n=6$, while the more heavily hole-doped $n=3$ compound remains nonsuperconducting \cite{Pan2026_science}. This series is especially useful because the layer number $n$ directly tunes the nominal Ni valence, $1+1/n$, without chemical substitution. It also provides a natural platform for studying layer differentiation, since inner and outer NiO$_2$ planes are structurally inequivalent. Recent theoretical phase-diagram analyses of Ruddlesden-Popper nickelates further indicate that local electronic correlations are among the key ingredients for superconductivity in this broader family \cite{Ouyang2025_prb_phase}.

The nonsuperconducting $n=3$ member raises a central question. Decreasing $n$ makes the electronic structure more two-dimensional and more cuprate-like in some aspects, and magnetic excitations persist into the overdoped nonsuperconducting regime \cite{Pan2026_science}. Nevertheless, superconductivity disappears at $n=3$. This suggests that cuprate-like band features or magnetic fluctuations alone may not be sufficient, and that the layer- and orbital-resolved Ni-$d$ correlations should be examined more directly. It also motivates a different tuning strategy. Instead of changing $n$, which changes the structure and Ni valence only in discrete steps, one may tune the carrier concentration chemically while keeping the low-$n$ framework. Along this materials-design direction, electron-doped Co-based La$_3$Ni$_2$O$_7$-like compounds were recently predicted as possible high-temperature superconducting candidates \cite{Wang2025_arxiv_co327}. A subsequent proposal showed that La$_4$Co$_2$NiO$_8$Cl$_2$ can realize a La$_4$Ni$_3$O$_{10}$-like crystal and strongly correlated electronic structure \cite{Jia2026_arxiv_co4310}. Recent DFT+DMFT work on electron versus hole doping in infinite-layer nickelates also emphasizes that electron compensation is not necessarily the simple inverse of hole doping because of rare-earth $5d$ self-doping effects \cite{DayRoberts2026_arxiv}. Spacer-layer Cl substitution is a natural candidate for this purpose. Replacing spacer-layer oxygen by Cl electron-dopes the system, can in principle continuously tune the Ni valence, and avoids direct substitution inside the NiO$_2$ planes. Since lower-$n$ members are structurally simpler and potentially easier to synthesize, such a route may provide experimentally accessible superconducting candidates.

In this work, we use charge self-consistent density functional theory combined with dynamical mean-field theory (DFT+DMFT) to study the layer- and orbital-resolved electronic correlations in multi-layer square-planar nickelates. We first investigate undoped La-based $n=3$--6 compounds and compare selected Nd-based systems to assess the role of rare-earth substitution. We then study Cl-doped $n=2$ and $n=3$ compounds, with the doping level chosen to match the nominal Ni valence of the optimally superconducting $n=6$ member. We find that the Ni-$d$ mass enhancement increases systematically from $n=3$ to $n=6$, and that inner NiO$_2$ layers are generally more strongly correlated than outer layers. The La and Nd compounds show very similar Ni-$d$ self-energies at the same $n$, indicating that the dominant trend is not controlled by rare-earth chemistry. Upon spacer-layer Cl doping, the low-$n$ compounds are driven into a correlation regime comparable to superconducting higher-$n$ nickelates, while the characteristic low-energy Ni-$d$ electronic structure is largely preserved. Computational details are summarized in Appendix~\ref{app:method}, and the La/Nd comparison is given in Appendix~\ref{app:rareearth}. These results suggest that spacer-layer electron compensation is a promising route to convert overdoped low-$n$ square-planar nickelates into experimentally testable superconducting candidates.

\section{Layer-resolved correlations in undoped La$_{n+1}$Ni$_n$O$_{2n+2}$}

\begin{figure}[t]
\centering
\includegraphics[width=\columnwidth]{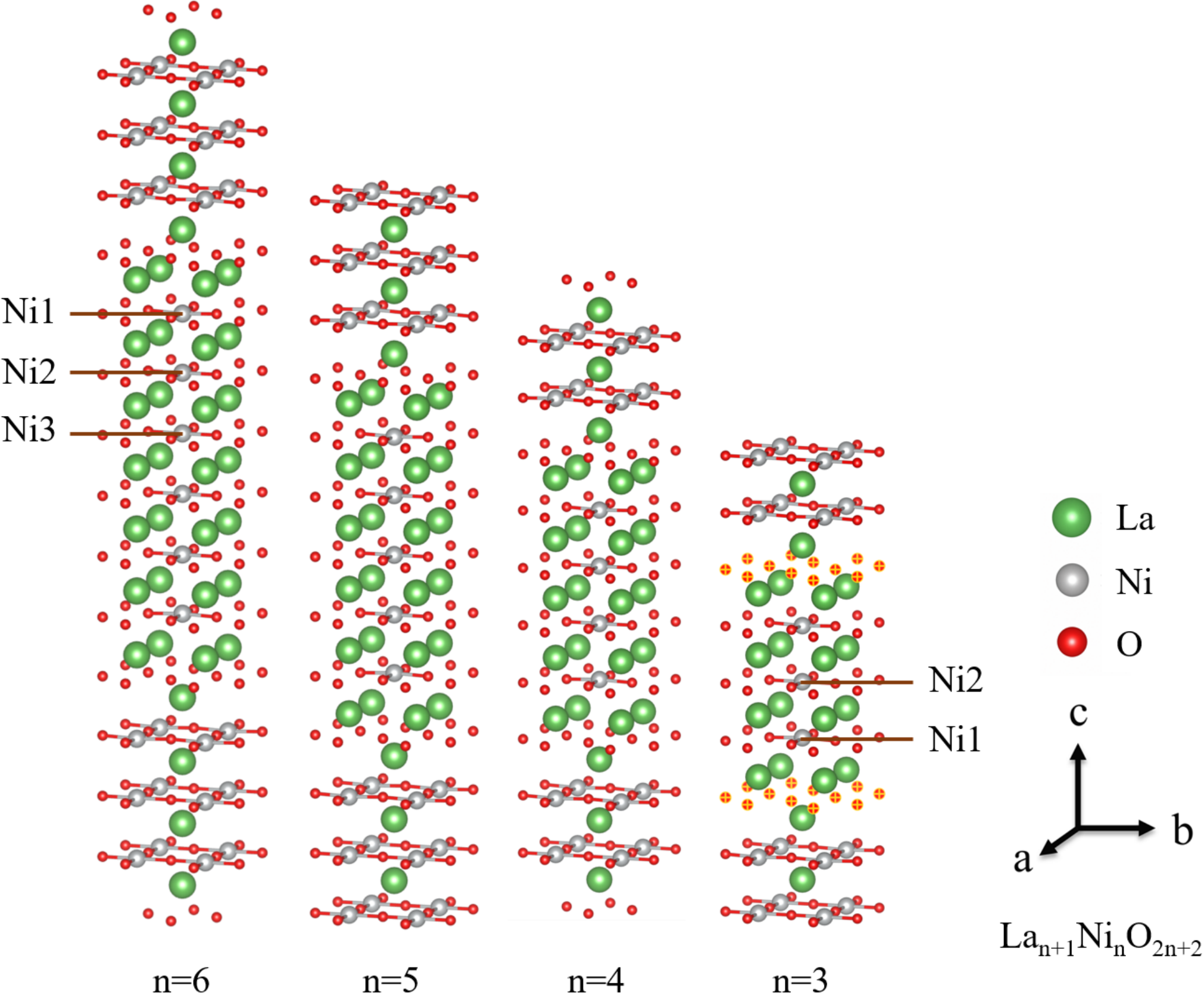}
\caption{
Crystal structures of undoped La$_{n+1}$Ni$_n$O$_{2n+2}$ with $n=3$, 4, 5, and 6.
The structures consist of square-planar NiO$_2$ blocks separated by spacer layers.
Representative inequivalent Ni sites are labeled as Ni1, Ni2, and Ni3 from the outer to the inner NiO$_2$ planes.
For the $n=3$ structure, the marked spacer-layer oxygen sites indicate the O sites used for O/Cl virtual-crystal substitution.
}
\label{fig:structures}
\end{figure}

\begin{figure}[t]
\centering
\includegraphics[width=\columnwidth]{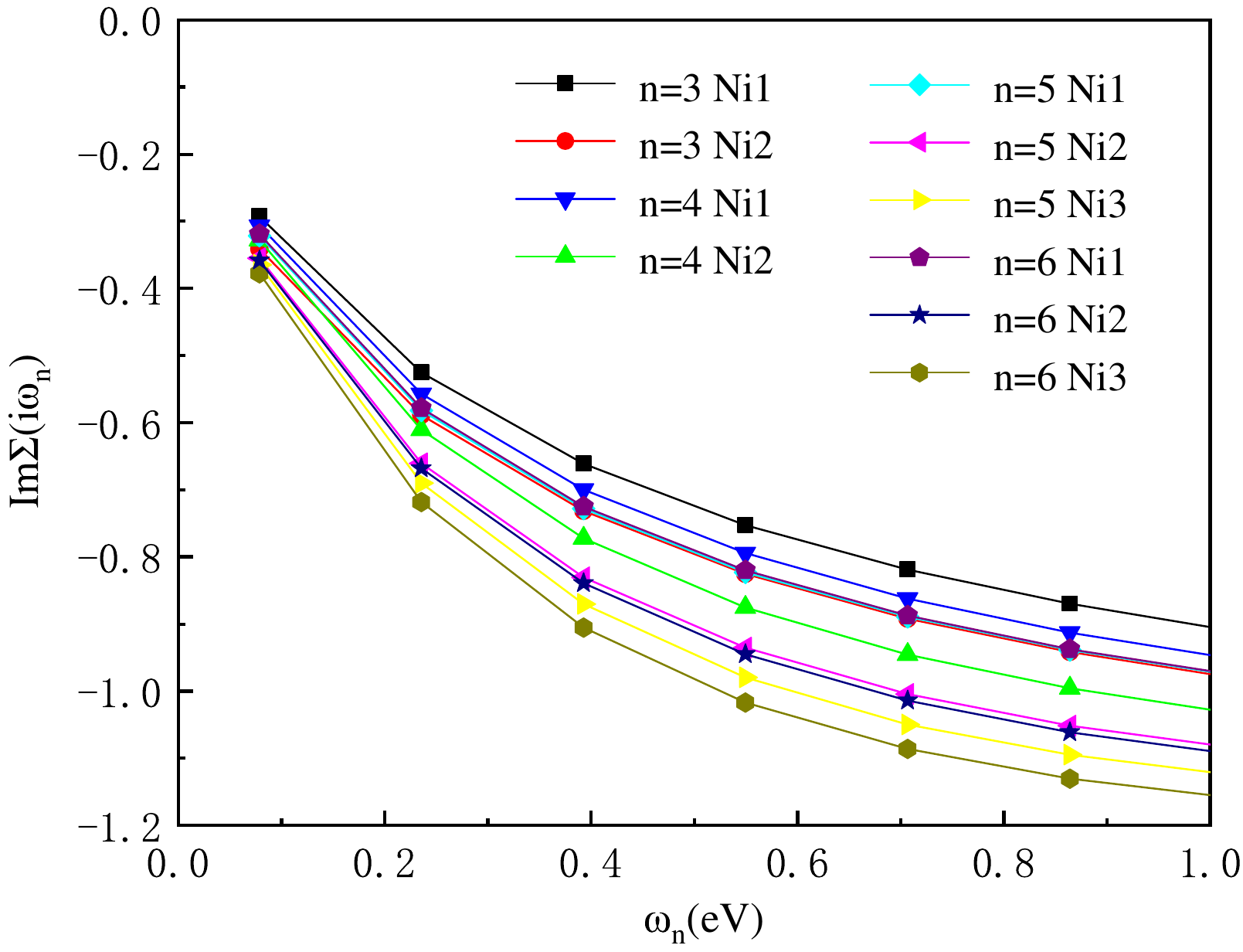}
\caption{
Imaginary part of the Matsubara self-energy, $\mathrm{Im}\Sigma(i\omega_n)$, for the Ni-$d_{x^2-y^2}$ orbital in undoped La$_{n+1}$Ni$_n$O$_{2n+2}$ with $n=3$--6.
Only inequivalent Ni sites are shown.
}
\label{fig:undoped_selfenergy}
\end{figure}

\begin{table}[t]
\caption{
Local orbital occupation number $N_d$ and effective mass enhancement $m^*/m$ of inequivalent Ni sites in the undoped La$_{n+1}$Ni$_n$O$_{2n+2}$ series with $n=3$--6 and in Cl-doped low-$n$ compounds.
The Cl-doped systems are VCA-averaged compounds with the average nominal Ni valence fixed to $+1.17$.
Ni1, Ni2, and Ni3 denote inequivalent Ni sites ordered from the outer to the inner NiO$_2$ planes.
}
\label{tab:undoped-la}
\centering
\footnotesize
\setlength{\tabcolsep}{2pt}
\begin{tabular}{@{}llcccc@{}}
\toprule
 & & \multicolumn{2}{c}{$N_d$} & \multicolumn{2}{c}{$m^*/m$} \\
\cmidrule(lr){3-4}\cmidrule(lr){5-6}
System & Ni site & $d_{z^2}$ & $d_{x^2-y^2}$ & $d_{z^2}$ & $d_{x^2-y^2}$ \\
\midrule
$n=3$ & Ni1 & 1.486 & 1.003 & 1.59 & 2.83 \\
 & Ni2 & 1.442 & 1.018 & 1.79 & 2.87 \\
\midrule
$n=4$ & Ni1 & 1.497 & 1.013 & 1.52 & 2.90 \\
 & Ni2 & 1.458 & 1.033 & 1.59 & 3.01 \\
\midrule
$n=5$ & Ni1 & 1.507 & 1.021 & 1.46 & 2.98 \\
 & Ni2 & 1.469 & 1.047 & 1.48 & 3.17 \\
 & Ni3 & 1.466 & 1.059 & 1.45 & 3.26 \\
\midrule
$n=6$ & Ni1 & 1.501 & 1.020 & 1.46 & 2.95 \\
 & Ni2 & 1.461 & 1.047 & 1.50 & 3.21 \\
 & Ni3 & 1.452 & 1.062 & 1.49 & 3.35 \\
\midrule
$n=2$ Cl-doped & Ni1 & 1.475 & 1.029 & 1.36 & 2.98 \\
\midrule
$n=3$ Cl-doped & Ni1 & 1.484 & 1.037 & 1.30 & 3.10 \\
 & Ni2 & 1.462 & 1.026 & 1.42 & 2.97 \\
\bottomrule
\end{tabular}
\end{table}

We begin by establishing the baseline electronic structure and layer-resolved correlation trends in the undoped La$_{n+1}$Ni$_n$O$_{2n+2}$ series with $n=3$--6. These compounds provide a direct reference for the experimentally synthesized multi-layer square-planar nickelates. The average nominal Ni valence is $1+1/n$, so that decreasing $n$ corresponds to increasing hole doping relative to the $d^9$ infinite-layer limit. The crystal structures are shown in Fig.~\ref{fig:structures}. Each unit cell contains two NiO$_2$ blocks separated by spacer layers. Within each block, inequivalent Ni sites can be classified according to their layer positions, with outer NiO$_2$ planes closer to the spacer layer and inner planes located deeper inside the block.

The DFT projected band structures of representative compounds are shown in Fig.~\ref{fig:bands_spectra}. In both the $n=3$ and $n=6$ systems, the low-energy electronic structure is dominated by Ni-$d_{x^2-y^2}$ states. The Ni-$d_{z^2}$ orbital also appears in the low-energy window, but it is much less correlated, as shown by the mass enhancements listed in Table~\ref{tab:undoped-la}. We therefore focus the main self-energy plots on the Ni-$d_{x^2-y^2}$ orbital, while keeping the full orbital-resolved occupancies and mass enhancements in the tables.

We now turn to the core many-body results. Figure~\ref{fig:undoped_selfenergy} shows $\mathrm{Im}\Sigma(i\omega_n)$ for the inequivalent Ni-$d_{x^2-y^2}$ orbital in the undoped La$_{n+1}$Ni$_n$O$_{2n+2}$ series. A clear systematic trend emerges as the layer number $n$ increases. The low-frequency magnitude of $\mathrm{Im}\Sigma(i\omega_n)$ becomes progressively more negative, and the slope near $\omega_n \to 0$ increases. This behavior indicates an enhancement of electronic correlations with increasing $n$. This trend is further quantified by the mass enhancement factors, estimated as
\begin{equation}
m^*/m \simeq 1 - \left. \frac{\partial \mathrm{Re}\Sigma(\omega)}{\partial \omega} \right|_{\omega \to 0}.
\end{equation}
As listed in Table~\ref{tab:undoped-la}, the largest $d_{x^2-y^2}$ mass enhancement increases from the relatively modest value of 2.87 in La$_4$Ni$_3$O$_8$ ($n=3$) to 3.35 in La$_7$Ni$_6$O$_{14}$ ($n=6$). This is consistent with the experimental phase diagram, where the $n=3$ compound is nonsuperconducting while superconducting signatures appear for higher-$n$ compounds, with the highest $T_c$ near $n=6$.

Furthermore, within each individual compound, we observe a robust layer-resolved correlation hierarchy. In the undoped series, the inner NiO$_2$ planes are generally more strongly correlated than the outer planes. For instance, in the $n=4$ compound, the inner-layer Ni has $m^*/m=3.01$ for the $d_{x^2-y^2}$ orbital, larger than the outer-layer value of 2.90. For $n=6$, the corresponding values are 3.35, 3.21, and 2.95 for the Ni3, Ni2, and Ni1 layers, respectively. This indicates that the inner layers, being further away from the spacer layers, host a more highly correlated electronic environment. The spacer-layer La$^{3+}$ charge reservoir provides holes primarily to the adjacent outer NiO$_2$ planes, leading to a higher hole concentration in the outer layers. This is consistent with the layer-resolved occupancies in Table~\ref{tab:undoped-la}.

A particularly important aspect of our results is the relationship between the Ni-$d_{x^2-y^2}$ orbital occupancy and the correlation strength. In a simple single-band Hubbard model, the correlation strength is maximized at half-filling, and doping reduces the correlation. As shown in Table~\ref{tab:undoped-la}, however, the $d_{x^2-y^2}$ occupancies in the $n=3$ compound are closest to one, while its correlations are weaker than those in the higher-$n$ compounds. This is not contradictory, because the effective low-energy degree of freedom is not a pure, isolated atomic Ni-$d_{x^2-y^2}$ orbital. It is an hybridized band formed by Ni-$d_{x^2-y^2}$ and O-$2p$ states. Due to the charge-transfer character, ligand holes on the oxygen sites can reduce the filling of the low-energy Ni-O hybridized band and drive it into the over-hole-doped regime. In this sense, the increase of the local Ni-$d_{x^2-y^2}$ occupancy with $n$ can still correspond to moving the effective low-energy band toward the correlated regime.

The DFT+DMFT spectral functions in Fig.~\ref{fig:bands_spectra} further show that the representative undoped systems remain correlated metals with renormalized Ni-$d$ bands near the Fermi level. The overall low-energy band structure is similar across the series, while the self-energy reveals the main distinction: the $n=3$ compound has weaker Ni-$d_{x^2-y^2}$ correlations than the superconducting higher-$n$ counterparts.

A comparison between La- and Nd-based compounds is given in Appendix~\ref{app:rareearth}, showing that rare-earth substitution does not substantially modify the Ni-$d$ correlations at fixed $n$.

\section{Spacer-layer Cl doping as electron compensation for low-$n$ nickelates}

\begin{figure}[t]
\centering
\includegraphics[width=\columnwidth]{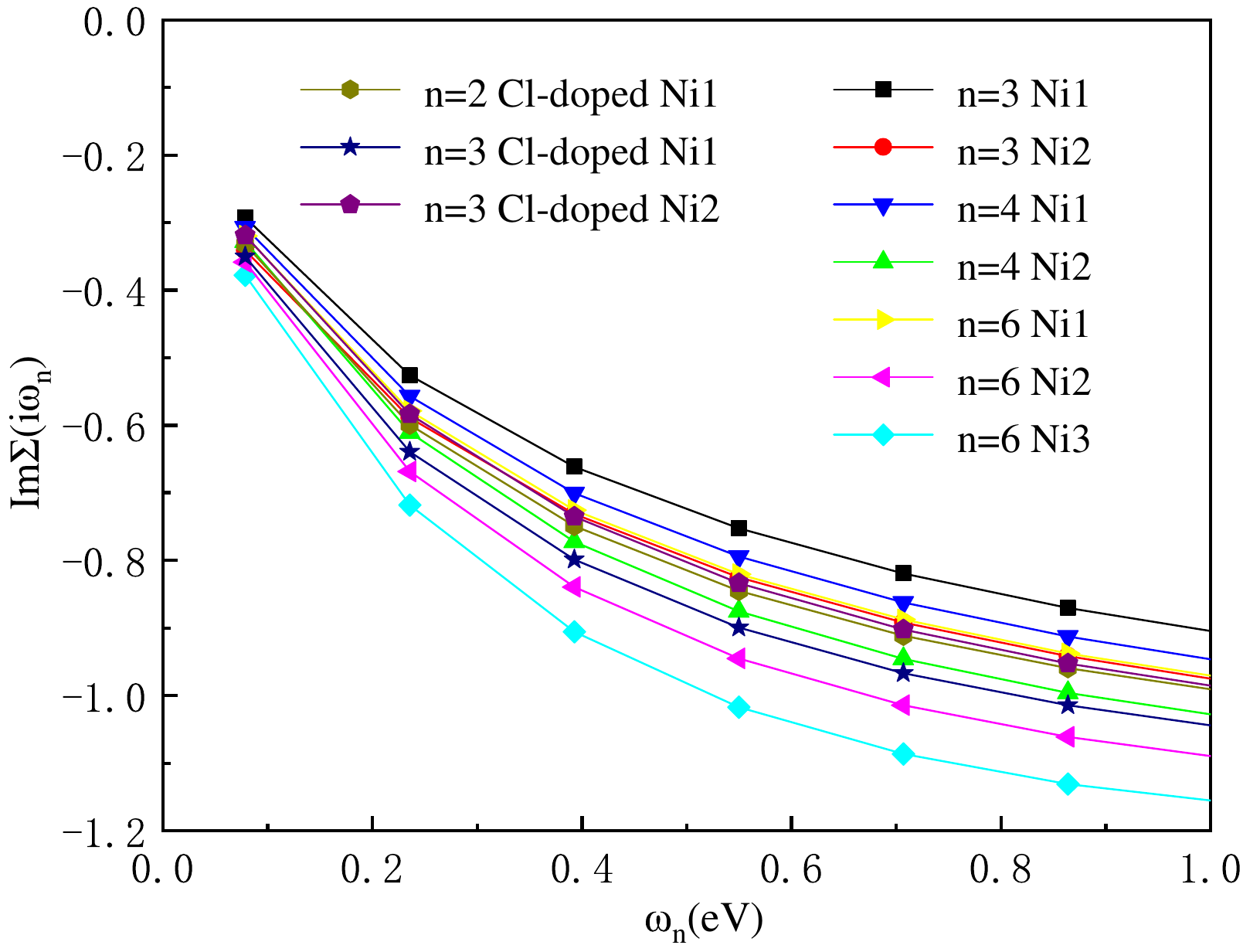}
\caption{
Comparison of $\mathrm{Im}\Sigma(i\omega_n)$ for the Ni-$d_{x^2-y^2}$ orbital in selected undoped La compounds and Cl-doped low-$n$ compounds.
The undoped references include La$_4$Ni$_3$O$_8$ ($n=3$), La$_5$Ni$_4$O$_{10}$ ($n=4$), and La$_7$Ni$_6$O$_{14}$ ($n=6$).
The Cl-doped compounds are treated within VCA and have the same average nominal Ni valence as the undoped $n=6$ compound.
}
\label{fig:cl_selfenergy}
\end{figure}

\begin{figure*}[t]
\centering
\includegraphics[width=\textwidth]{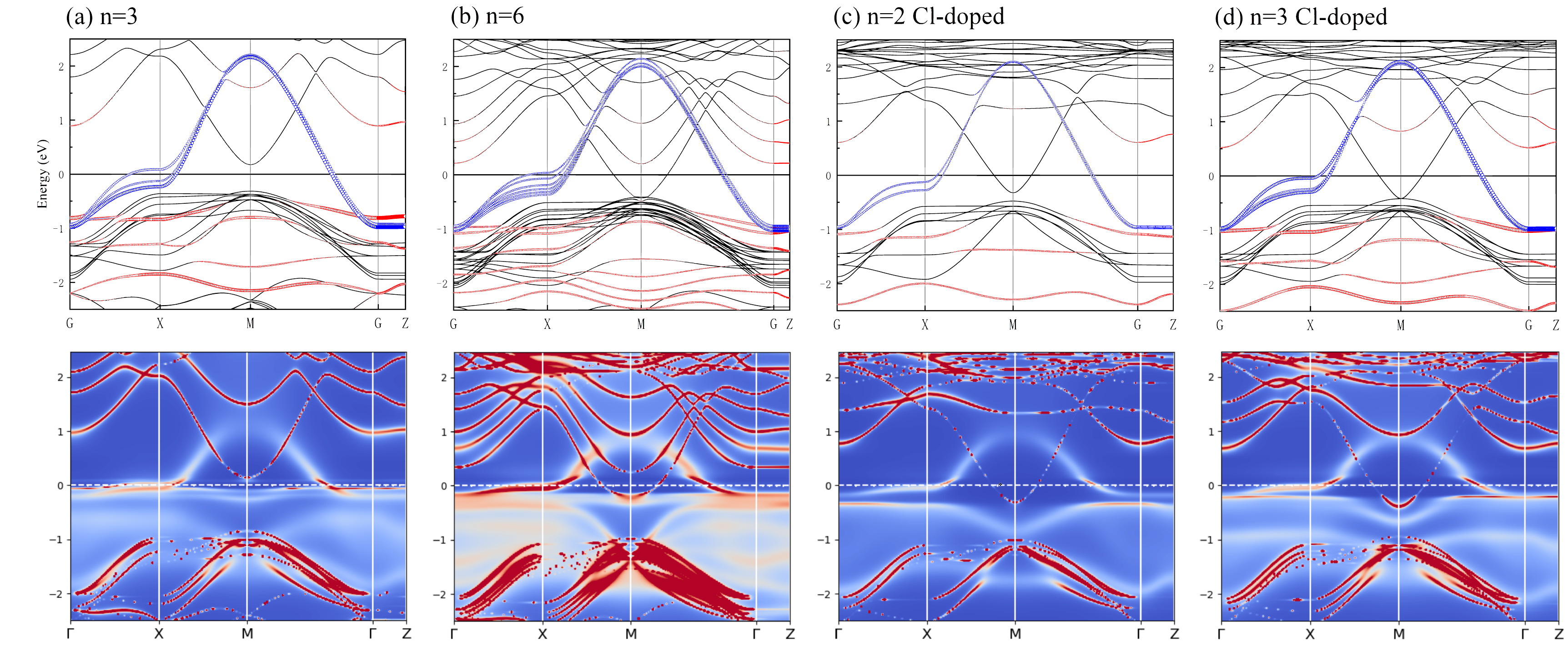}
\caption{
DFT projected band structures and DFT+DMFT spectral functions of representative compounds.
The four systems are undoped La$_4$Ni$_3$O$_8$ ($n=3$), undoped La$_7$Ni$_6$O$_{14}$ ($n=6$), Cl-doped La$_3$Ni$_2$O$_{5.33}$Cl$_{0.67}$, and Cl-doped La$_4$Ni$_3$O$_{7.50}$Cl$_{0.50}$.
The projected band structures highlight the Ni-$d_{z^2}$ (red) and Ni-$d_{x^2-y^2}$ (blue) orbital characters.
}
\label{fig:bands_spectra}
\end{figure*}

We next examine whether electron compensation can drive low-$n$ square-planar nickelates into the correlation regime of superconducting higher-$n$ compounds. In the undoped series, the $n=2$ and $n=3$ compounds are more strongly hole-doped than the $n=6$ compound. We therefore introduce Cl substitution on the spacer-layer oxygen sites using the virtual crystal approximation. The Cl concentration is chosen such that the average nominal Ni valence becomes $+1.17$, the same as in the $n=6$ compound. This corresponds to La$_3$Ni$_2$O$_{5.33}$Cl$_{0.67}$ for the $n=2$ compound and La$_4$Ni$_3$O$_{7.50}$Cl$_{0.50}$ for the $n=3$ compound in the averaged composition. The substitution is restricted to spacer-layer oxygen sites, leaving the NiO$_2$ planes intact and preserving the original $I4/mmm$ symmetry within VCA.

We now evaluate the electronic correlations of these Cl-doped compounds. Figure~\ref{fig:cl_selfenergy} compares the $d_{x^2-y^2}$ self-energies of the Cl-doped low-$n$ compounds with those of selected undoped reference compounds. The Cl-doped $n=2$ compound shows a much stronger self-energy than the undoped $n=3$ compound. Its $d_{x^2-y^2}$ mass enhancement is 2.98, close to the inner-layer value of undoped La$_5$Ni$_4$O$_{10}$ ($n=4$), as listed in Table~\ref{tab:undoped-la}. Thus, electron compensation brings the $n=2$ compound into the same correlation range as the experimentally superconducting higher-$n$ region.

The effect is more layer selective in the Cl-doped $n=3$ compound. The outer-layer Ni-$d_{x^2-y^2}$ self-energy is strongly enhanced, with $m^*/m=3.10$, larger than in the undoped $n=3$ compound and comparable to the undoped $n=5$--6 compounds. In contrast, the inner-layer value is 2.97, only moderately changed from the undoped case. As a result, the usual hierarchy of stronger inner-layer correlations is reversed in the Cl-doped $n=3$ compound, where the outer layer becomes more strongly correlated than the inner layer. This layer selectivity is consistent with the spacer-layer location of the Cl substitution, which most directly affects the adjacent outer NiO$_2$ planes.

\begin{figure*}[t]
\centering
\includegraphics[width=\textwidth]{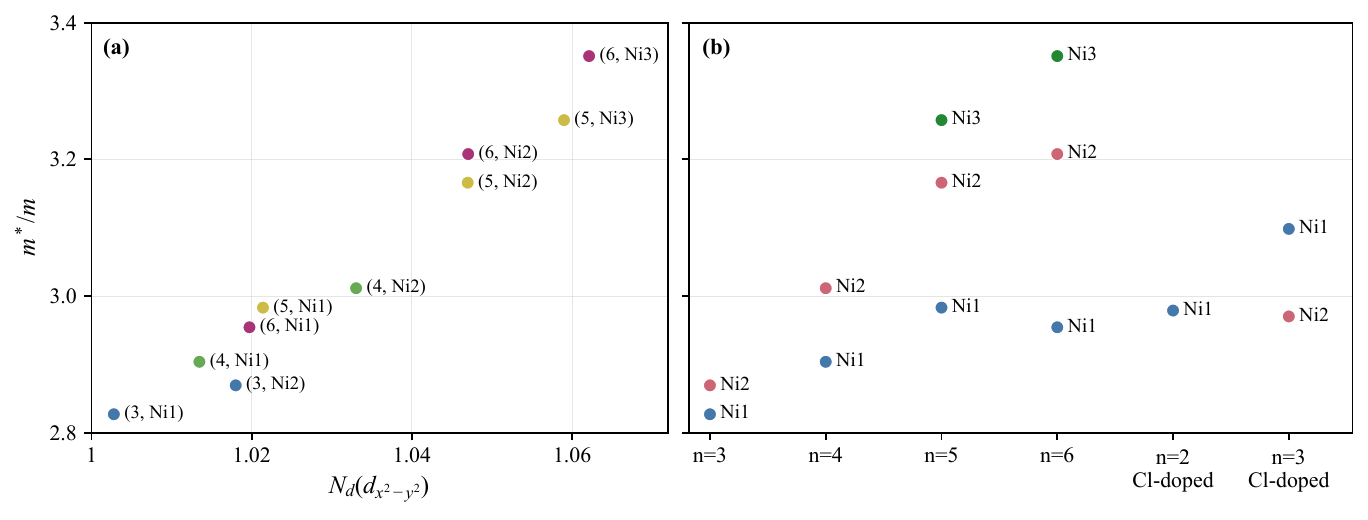}
\caption{
Summary of the layer-resolved Ni-$d_{x^2-y^2}$ mass enhancement.
(a) Relation between the local Ni-$d_{x^2-y^2}$ occupation $N_d(d_{x^2-y^2})$ and the effective mass enhancement $m^*/m$ for inequivalent Ni sites in undoped La$_{n+1}$Ni$_n$O$_{2n+2}$ with $n=3$--6.
The labels denote $(n,\mathrm{Ni}m)$, where Ni1, Ni2, and Ni3 are ordered from the outer to the inner NiO$_2$ planes.
(b) Comparison of $m^*/m$ across selected undoped and Cl-doped systems.
The Cl-doped $n=2$ and $n=3$ compounds are treated within VCA and have the same average nominal Ni valence as the undoped $n=6$ compound.
}
\label{fig:mstar_summary}
\end{figure*}

The mass-enhancement trends are summarized in Fig.~\ref{fig:mstar_summary}. In the undoped series, the Ni-$d_{x^2-y^2}$ mass enhancement increases with the local $d_{x^2-y^2}$ occupancy and with increasing layer number. When the Cl-doped compounds are placed on the same scale, the doped $n=2$ and $n=3$ systems move into the correlation range of the superconducting higher-$n$ compounds rather than that of the weakly correlated undoped $n=3$ parent.

The projected band structures and DFT+DMFT spectral functions of the Cl-doped compounds are shown in Fig.~\ref{fig:bands_spectra}. In both doped systems, the low-energy bands remain dominated by Ni-$d_{x^2-y^2}$ character. No additional low-energy impurity band appears within the VCA treatment. The spectral functions also retain the correlated metallic features seen in the undoped compounds, with renormalized Ni-$d$ bands near the Fermi level. Therefore, spacer-layer Cl doping does not create a qualitatively different low-energy electronic structure. Instead, it tunes the carrier concentration and enhances the Ni-$d_{x^2-y^2}$ correlations within the same square-planar nickelate platform.

These results support the following picture. The undoped $n=3$ compound is on the over-hole-doped, weaker-correlation side of the multi-layer nickelate phase diagram. Spacer-layer Cl substitution compensates this over-hole doping and moves the low-$n$ compounds toward the correlation regime of superconducting higher-$n$ nickelates. The Cl-doped $n=2$ and $n=3$ square-planar nickelates are therefore promising superconducting candidates for experimental exploration.

\section{Discussion and Conclusion}

To place our findings in a broader context, we return to the central puzzle posed by the experimental phase diagram of multi-layer square-planar nickelates $\mathrm{R}_{n+1}\mathrm{Ni}_n\mathrm{O}_{2n+2}$. As established in recent studies, this family serves as a unique platform where the number of NiO$_2$ layers $n$ acts as a structural tuning parameter. While superconductivity has been observed in the $n=4$--8 members, with the transition temperature $T_c$ peaking near $n=6$, the $n=3$ member, $\mathrm{R}_4\mathrm{Ni}_3\mathrm{O}_8$, remains nonsuperconducting. This observation is particularly intriguing because lower-layer systems are structurally closer to cuprates and might be expected to host stronger cuprate-like physics. However, the lower dimensionality or structural similarity to cuprates alone is not a sufficient condition for superconductivity. Instead, our layer-resolved Ni-$d$ electronic correlations provide a more direct microscopic criterion. The emergence of superconductivity in these materials requires the low-energy electronic states to fall within a strongly correlated metallic window, which the heavily hole-overdoped $n=3$ compound fails to reach.

Our work advances the understanding of these materials in three major ways. First, we have pushed the conventional nominal-valence analysis, which typically relies on the average Ni nominal valence $1+1/n$, to the layer- and orbital-resolved levels. Our calculations reveal that the average valence is an insufficient descriptor of the low-energy physics because of the pronounced spatial anisotropy within the NiO$_2$ blocks. The inequivalent planes exhibit distinct $d_{x^2-y^2}$ self-energies and mass enhancement factors. This layer-resolved correlation hierarchy highlights that the electronic properties of multi-layer nickelates should be understood as a spatially modulated correlated state rather than treated as a homogeneous average.

Second, we have demonstrated that La-based calculations provide a computationally cleaner theoretical proxy for the Nd-based experimental systems. As detailed in Appendix~\ref{app:rareearth}, by performing a systematic comparison between the La and Nd analogues for $n=3$ and $n=4$, we find that the low-energy Ni-$d$ self-energies, occupancies, and mass enhancements are nearly identical. This indicates that rare-earth substitution is not a dominant variable governing the local correlation strength within the Ni-$d$ manifold. This comparison supports our subsequent material design based on the virtual crystal approximation (VCA) within the La-based framework and makes the predictions relevant to Nd-based experimental compounds.

Third, we have translated the discrete structural tuning of changing the layer number $n$ into a more continuously tunable chemical strategy based on spacer-layer electron compensation. In the experimental phase diagram, $n$ is restricted to integer values, and varying $n$ simultaneously alters the crystal structure, the number of layers, the interlayer coupling, and the nominal filling. In contrast, spacer-layer O/Cl substitution provides a more direct tuning knob. By replacing O$^{2-}$ with Cl$^{-}$ only on the spacer-layer oxygen sites, electrons are compensated into the system to tune the Ni valence while keeping the active NiO$_2$ planes structurally intact and free from direct chemical substitution.

From an experimental perspective, this spacer-layer Cl-doping route offers several advantages. Low-layer systems such as $n=2$ and $n=3$ possess simpler crystal structures, which may make them easier to synthesize than their high-$n$ counterparts. The ability to tune the O/Cl ratio can allow systematic mapping of the correlation landscape and help locate the optimal doping level beyond the discrete limitations of the parent compounds. Our finding that Cl-doped $n=2$ and $n=3$ compounds exhibit Ni-$d_{x^2-y^2}$ mass enhancements close to those of superconducting reference compounds makes them concrete and promising candidates for experimental realization.

To maintain a rigorous perspective, we must acknowledge the limitations of our theoretical approach. While DFT+DMFT self-energies and mass enhancements provide a useful measure of local electronic correlations, they do not constitute direct proof of superconductivity. Our results show that spacer-layer doping positions the low-$n$ candidates within the correlated metallic window associated with known superconductors, but the actual transition to a superconducting state may depend on additional factors. Since our calculations focus on local self-energies, we do not compute momentum-dependent magnetic fluctuations or pairing eigenvalues, leaving the detailed pairing mechanism open for future investigation.

Nevertheless, our predictions offer clear and testable pathways for experimental verification. We propose the synthesis of Cl-doped low-layer square-planar nickelates, such as $\mathrm{La}_3\mathrm{Ni}_2\mathrm{O}_{5.33}\mathrm{Cl}_{0.67}$ and $\mathrm{La}_4\mathrm{Ni}_3\mathrm{O}_{7.50}\mathrm{Cl}_{0.50}$, via topotactic reduction of suitable precursor phases. Recent progress in controllable topotactic reduction methods suggests that such chemical routes are becoming increasingly accessible for nickelate materials \cite{Zhang2025_commmater_topotactic}. Angle-resolved photoemission spectroscopy can be employed to map the low-energy Ni-$d$ bands and verify the shift of the Fermi level. X-ray absorption spectroscopy and electron energy-loss spectroscopy can be used to examine the Ni valence and probe the charge distribution. Resonant inelastic x-ray scattering can further measure the spin excitation spectra and determine whether the magnetic fluctuations in these Cl-doped systems resemble those observed in superconducting high-$n$ compounds.

In conclusion, we have performed a systematic charge self-consistent DFT+DMFT study of the electronic correlations in multi-layer square-planar nickelates. Our analysis spans the undoped $\mathrm{La}_{n+1}\mathrm{Ni}_n\mathrm{O}_{2n+2}$ series with $n=3$--6, a comparison between La- and Nd-based analogues for $n=3$ and $n=4$, and a material design based on spacer-layer Cl-doped $n=2$ and $n=3$ systems. These results identify spacer-layer electron compensation as a promising route for designing low-$n$ square-planar nickelate superconductors.

\begin{acknowledgments}
This work was supported by the National Key R\&D Program of China (Grants No. 2024YFA1408602 and No. 2024YFA1408601) and the National Natural Science Foundation of China (Grant No. 12434009). Z.-Y.L. was also supported by the Innovation Program for Quantum Science and Technology (Grant No. 2021ZD0302402). Computational resources were provided by the Physical Laboratory of High Performance Computing in Renmin University of China.
\end{acknowledgments}

\appendix

\section{Method}
\label{app:method}

Our calculations were performed using density functional theory combined with dynamical mean-field theory (DFT+DMFT). The nonmagnetic DFT calculations were carried out within the full-potential linearized augmented-plane-wave method as implemented in the \textsc{wien2k} package \cite{Blaha2020_jcp}. The charge self-consistent DFT+DMFT calculations were performed using the eDMFT implementation \cite{Haule2010_prb}. All calculations were done in the paramagnetic state at $T=290$ K. The Ni-$3d$ $e_g$ orbitals, including $d_{z^2}$ and $d_{x^2-y^2}$, were treated as the correlated subspace with $U=5.0$ eV and $J=1.0$ eV. For inequivalent Ni atoms, independent impurity problems were constructed and solved. Projectors onto the correlated orbitals were constructed using an energy window from $-10$ to $10$ eV with respect to the Fermi level, which includes the full O-$2p$ manifold. The quantum impurity problems were solved by the continuous-time quantum Monte Carlo method in the hybridization-expansion formulation \cite{Haule2007_prb}. The double-counting correction was treated using the exact double-counting scheme \cite{Haule2015_prl}. Real-frequency self-energies and momentum-resolved spectral functions were obtained by the maximum entropy method \cite{Jarrell1996_physrep}.

To simulate Cl substitution in the low-$n$ compounds, we used the virtual crystal approximation (VCA). In the $n=2$ and $n=3$ systems, Cl was introduced on the spacer-layer oxygen sites, which are located between neighboring NiO$_2$ blocks rather than inside the NiO$_2$ planes. The O/Cl virtual atom was constructed to represent a uniform substitution on these spacer-layer sites. The Cl concentration was chosen such that the average nominal Ni valence becomes $+1.17$, the same as that of the $n=6$ compound. This corresponds to electron compensation of the over-hole-doped low-$n$ systems. Within this VCA treatment, the original $I4/mmm$ symmetry was retained, representing a spatially averaged, symmetry-preserving doped structure.

For the analysis of electronic correlations, we extracted the orbital-resolved self-energies, Ni-$d$ occupancies, and quasiparticle mass enhancements for each inequivalent Ni site. The mass enhancement was estimated from the low-frequency behavior of the self-energy. To compare the doped low-$n$ compounds with the undoped higher-$n$ series, all quantities were analyzed in a layer- and orbital-resolved manner.

\section{The effect of rare-earth substitution on Ni-$d$ correlations}
\label{app:rareearth}

\begin{figure}[t]
\centering
\includegraphics[width=\columnwidth]{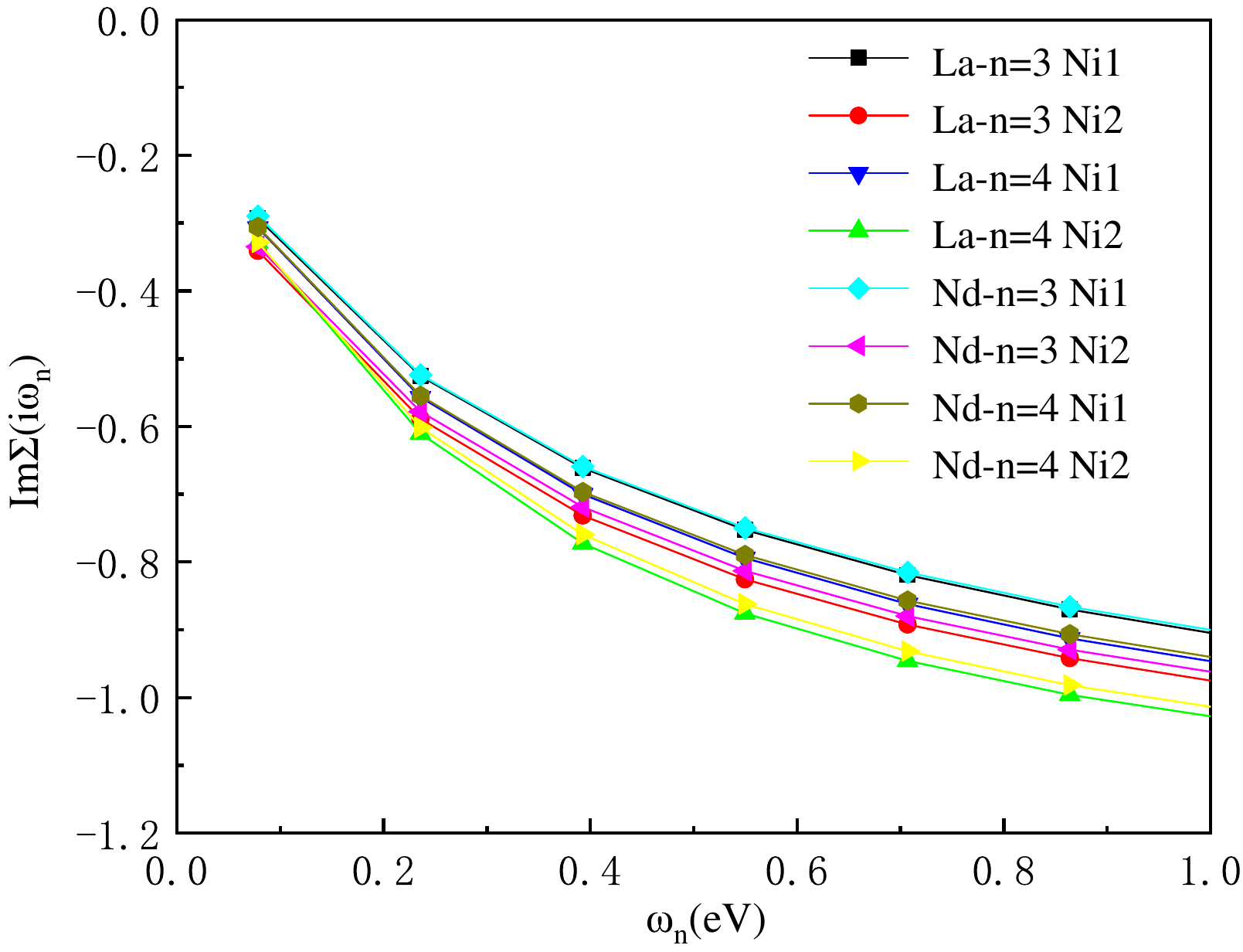}
\caption{
Comparison of $\mathrm{Im}\Sigma(i\omega_n)$ for the Ni-$d_{x^2-y^2}$ orbital in La- and Nd-based $n=3$ and $n=4$ compounds.
The corresponding compounds are La$_4$Ni$_3$O$_8$, Nd$_4$Ni$_3$O$_8$, La$_5$Ni$_4$O$_{10}$, and Nd$_5$Ni$_4$O$_{10}$.
}
\label{fig:rareearth_selfenergy}
\end{figure}

\begin{table}[t]
\caption{
Local orbital occupation number $N_d$ and effective mass enhancement $m^*/m$ of inequivalent Ni sites in La- and Nd-based $n=3$ and $n=4$ compounds.
Ni1 and Ni2 denote inequivalent Ni sites ordered from the outer to the inner NiO$_2$ planes.
}
\label{tab:rareearth}
\centering
\footnotesize
\setlength{\tabcolsep}{2pt}
\begin{tabular}{@{}llcccc@{}}
\toprule
 & & \multicolumn{2}{c}{$N_d$} & \multicolumn{2}{c}{$m^*/m$} \\
\cmidrule(lr){3-4}\cmidrule(lr){5-6}
System & Ni site & $d_{z^2}$ & $d_{x^2-y^2}$ & $d_{z^2}$ & $d_{x^2-y^2}$ \\
\midrule
La, $n=3$ & Ni1 & 1.486 & 1.003 & 1.59 & 2.83 \\
 & Ni2 & 1.442 & 1.018 & 1.79 & 2.87 \\
\midrule
Nd, $n=3$ & Ni1 & 1.489 & 1.004 & 1.57 & 2.87 \\
 & Ni2 & 1.443 & 1.017 & 1.79 & 2.89 \\
\midrule
La, $n=4$ & Ni1 & 1.497 & 1.013 & 1.52 & 2.90 \\
 & Ni2 & 1.458 & 1.033 & 1.59 & 3.01 \\
\midrule
Nd, $n=4$ & Ni1 & 1.501 & 1.015 & 1.50 & 2.96 \\
 & Ni2 & 1.459 & 1.033 & 1.57 & 3.00 \\
\bottomrule
\end{tabular}
\end{table}

The experimental multi-layer square-planar nickelates are Nd-based, whereas most calculations in this work are performed for La-based compounds. We therefore compare La and Nd compounds at the same layer number to examine whether rare-earth substitution qualitatively changes the Ni-$d$ correlations.

Figure~\ref{fig:rareearth_selfenergy} shows the $d_{x^2-y^2}$ Matsubara self-energies for La$_4$Ni$_3$O$_8$, Nd$_4$Ni$_3$O$_8$, La$_5$Ni$_4$O$_{10}$, and Nd$_5$Ni$_4$O$_{10}$. For both $n=3$ and $n=4$, the corresponding La and Nd self-energies are very close. The same conclusion is obtained from the mass enhancements and orbital occupancies in Table~\ref{tab:rareearth}. In the $n=3$ compounds, the inner-layer $d_{x^2-y^2}$ mass enhancement is 2.87 for La and 2.89 for Nd. In the $n=4$ compounds, the corresponding inner-layer values are 3.01 and 3.00. The outer-layer values show similarly small differences.

These results indicate that replacing La by Nd does not substantially modify the low-energy Ni-$d$ self-energies at fixed $n$. The dominant trend discussed above is therefore controlled mainly by the NiO$_2$ electronic structure and the layer-dependent filling, rather than by the rare-earth ion.

\bibliography{n-layer-LaNiO2}

\end{document}